\documentclass[a4paper,referee,usenatbib]{mnras}
\usepackage{newtxtext,newtxmath}
\usepackage[dvipdfmx]{graphicx}
\usepackage{mediabb}

\usepackage[T1]{fontenc}
\usepackage{ae,aecompl}

\usepackage{graphicx}     
\usepackage{amsmath}     
\usepackage{amssymb}     
\usepackage{longtable} 
\usepackage{pdflscape}

\title[A high-B PSR survey with {\it Swift}]{A high-magnetic-field radio pulsar survey with {\it Swift}/{\rm XRT} }

\author[Watanabe et al.]{
Eri Watanabe,$^{1}$\thanks{E-mail: eri@ksirius.kj.yamagata-u.ac.jp}
Shinpei Shibata,$^{2}$
Takanori Sakamoto,$^{3}$
Aya Bamba$^{4,5}$
\\
$^{1}$\scriptsize{School of Science and Technology, Yamagata University, 1-4-12 Kojirakawa, Yamagata, Japan}\\
$^{2}$\scriptsize{Department of Physics, Yamagata University, 1-4-12 Kojirakawa, Yamagata, Japan}\\
$^{3}$\scriptsize{Department of Physics and Mathematics, Aoyama Gakuin University, 5-10-1 Fuchinobe, Chuo-ku, Sagamihara, Kanagawa 252-5258, Japan}\\
$^{4}$\scriptsize{Department of Physics, The University of Tokyo, 7-3-1 Hongo, Bunkyo-ku, Tokyo 113-0033, Japan}\\
$^{5}$\scriptsize{Research Center for the Early Universe, School of Science, The University of Tokyo, 7-3-1 Hongo, Bunkyo-ku, Tokyo 113-0033, Japan}}

\date{Accepted XXX. Received YYY; in original form 21 August 2018}

\pubyear{2018}

\begin{document}
\label{firstpage}
\pagerange{\pageref{firstpage}--\pageref{lastpage}}
\maketitle

\begin{abstract}
We present the X-ray survey results of high-magnetic-field radio pulsars (high-B PSRs) with {\it Swift}/{\rm XRT}.
X-ray observations of the rotation-powered pulsars with the dipole magnetic field ${\it B_{\rm d}}$ near the quantum critical field ${\it B_{\rm q}}=4.4\times10^{13}$~G is of great importance for understanding the transition between the rotation-powered pulsars and the magnetars, because there are a few objects that have magnetar-like properties.
Out of the 27 high-B PSRs that are in the ATNF pulsar catalogue but have not been reported or have no effective upper-limits in the X-ray bands, we analyze the {\it Swift}/{\rm XRT} data for 21 objects, where 6 objects are newly observed and 15 objects are taken from the archival data.
As a result, we have new $3\sigma$ upper-limits for all the 21 objects. 
Since the upper-limits are tight, we conclude that we do not find any magnetar-like high-B PSRs such as PSR~J1819$-$1458. 
The probability of the high X-ray efficiency in the high-B PSRs is obtained to be $11\%-29\%$.
Combining the previous observations, we discuss which parameter causes magnetar-like properties. It may be suggested that the magnetar-like properties appear only when ${\it B_{\rm d}}\gtrsim10^{13.5}$~G for the radio pulsar population. This is true even if the radio-quiet high-B RPP are included.
\end{abstract}



\clearpage
\section{Introduction}
Isolated neutron stars are known to have two major populations 
according to the source of their energy.
One is the rotation-powered pulsars (RPPs)
(for a review, see \citealt{2015SSRv..191..207B}).
These are mostly observed as radio pulsars (PSRs) or gamma-ray pulsars.
The other population is the high magnetic pulsars, called magnetars for short
(for a review, see \citealt{kaspi2017} and \citealt{2015RPPh...78k6901T}).
These are characterised by frequent bursting activity, lasting from a fraction of a second to several tens of seconds, and by high X-ray luminosity larger than the spin-down luminosity $L_{\rm rot}$.

What causes this distinctive manifestation of neutron stars as either 
RPPs or magnetars is not yet understood.
The dipole component of the neutron star magnetic field
can be inferred from the spin period $P$ and its time derivative $\dot{P}$ as
$B_d \approx 1.1\times 10^{12} (P/1 \mbox{s})^{1/2} (\dot{P}/10^{-15})^{1/2}$~G. 
Since most magnetars have a very strong dipole magnetic field 
exceeding the quantum critical field 
${B_{\rm q}} = m^2 c^3/\hbar e =4.4 \times10^{13}$~G,
it was thought that the strength of the dipole field is the key parameter.
However, discovery of the ``weak-field magnetar'' SGR~0418$+$5729, 
whose dipole field
$B_d \sim 6.1 \times 10^{12}$~G is much less than $B_{\rm q}$
\citep{2010MNRAS.405.1787E, Rea2010, 2013ApJ...770...65R}, indicates that
dipole field strength is not always essential.
Theoretically, the toroidal field is proposed to play an essential role
for magnetar activity \citep{2011ApJ...741..123P}.
The toroidal field together with the poloidal field evolves interactively 
from very
large values $\gtrsim 10^{14}$~G
at birth to their present state,
and is expected to form multi-pole fields
that cause bursts and extra heating.

A small group of RPPs called high-magnetic-field RPPs (high-B RPPs), some of which are radio-quiet pulsars, may 
be key to understand the strong magnetic fields of the neutron stars.
The inferred dipole magnetic field of high-B RPPs
is similar to or larger than ${B_{\rm q}}$.
These are interestingly symbiotic objects in the sense that
they exhibit both RPP-like and magnetar-like features.
In fact, high-B RPPs generally behave very much like RPPs, but
two recently exhibited magnetar-like outbursts 
(see \citealt{Archibald2016} regarding an outburst from PSR~J1119$-$6127,
\citealt{2008Sci...319.1802G} for PSR~J1846$-$258),
and three are known to have high X-ray luminosity as compared 
with the typical luminosity of RPPs, $\sim 10^{-3} L_{\rm rot}$ 
\citep{1988ApJ...332..199S, 1997A&A...326..682B, Becker2009, Kargaltsev2012, shibata2016}.

To provide better and possibly new insights for the symbiosis seen 
in high-B RPPs, 
we intend to sample more X-ray counterparts of high-B RPPs.
Previously, some high-B RPPs were observed intensively 
\citep{Olausen2013}, and
an X-ray counterpart survey with {\it Chandra} and {\it XMM-Newton} 
was done by \citet{Prinz2015}.
It is notable here that the magnetospheric emission of RPPs declines rapidly with spin-down,
so that thermal radiation, which may be due to strong multi-pole magnetic fields, is easier to be detected
in objects with $L_{\rm rot} \lesssim 10^{35} \mbox{erg s$^{-1}$}$.
However, the survey for such objects with relatively small spin-down luminosity 
is not complete.
Toward a more complete survey, in this paper we report the first results of 
a systematic survey for X-ray counterparts
using archival data and additional observations 
with the X-ray Telescope ({\rm XRT}; \citealt{2005SSRv..120..165B}) 
onboard {\it the Neil Gehrels Swift Observatory} \citep{2004ApJ...611.1005G}.

\section{Data Preparation and Observation}

Using the Australia Telescope National Facility (ATNF) pulsar catalogue
\footnote{version 1.55, http://www.atnf.csiro.au/research/pulsar/psrcat/expert.html} 
\citep{2005AJ....129.1993M}, 
we made a list of high-B PSRs with inferred dipole magnetic fields 
above $10^{13}$~G.
Since our targets are PSRs, 
we excluded magnetars, X-ray isolated neutron star (XINS), radio-quiet pulsars,
 and binary members from the ATNF catalogue using type codes AXP, XINS, NRAD 
and BINARY \footnote{The type codes AXP, XINS, NRAD, and BINARY 
respectively indicate an anomalous X-ray pulsar or soft gamma-ray repeater, 
an isolated neutron star with pulsed thermal X-ray emission 
but no detectable radio emission, 
a spin-powered pulsar with pulsed emission only at infrared or 
higher frequencies, or a binary system.}.
A total of 56 objects were selected.
After a literature search, 27 objects were found to have no X-ray detections 
or no effective upper-limits less than 
$10^{35}$~erg~s$^{-1}$.
We searched {\it Swift}/{\rm XRT} 
archival data that include the radio positions of these 27 high-B PSRs 
within 12 arcmin from the {\it Swift} pointing position in photon 
counting (PC) mode.
As a result, available data were found for 16 of the 27 objects, 
as listed in Table~\ref{tab:arc}.
There were no X-ray observations for the remaining 11 objects.

We attempted additional {\it Swift} observations for the remaining objects.
We selected objects that can be detected at 10 sigma or more with 2~ks 
exposure time if their luminosities are as large as that of 
PSR~J1819$-$1458 ($1.75\times10^{33}$~erg~s$^{-1}$; \citealt{McLaughlin2007}), 
which is considered to be a typical value of magnetic origin for the thermal 
radiation seen in high-B PSRs.
As a result, we selected 5 objects.
The observation parameters for these objects are given in Table~\ref{tab:obs}.
Note that we performed a second observation for PSR J1851$+$0118, since it gave 4$\sigma$ detection in the first analysis of the archival data (Table~\ref{tab:obs}).

\begin{figure}
\begin{center}
\includegraphics[width=0.6\columnwidth]{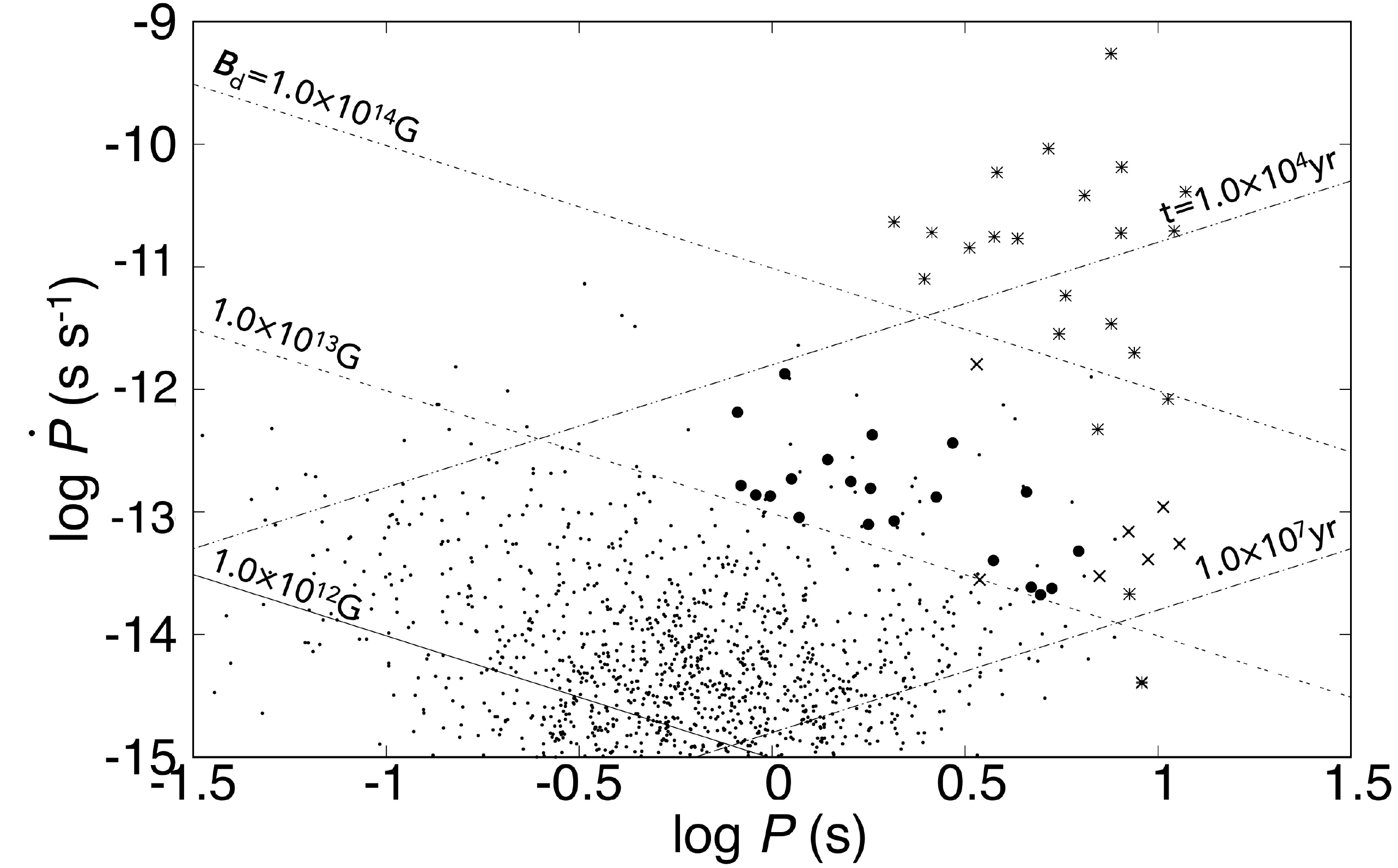}
\caption{$P-\dot{P}$ diagram for our samples, indicated by filled black circles with other populations.
Dots indicate RPPs listed in the ATNF pulsar catalogue. 
$\times$ symbols and asterisks indicate XINSs \citep{vigano2013} and magnetars \citep{2014ApJS..212....6O}, respectively.}
\label{PPdot}
\end{center}
\end{figure}

We performed the analysis for the 21 pulsars. 
The locations in the $P-\dot{P}$ diagram of those pulsars are shown 
in Figure~\ref{PPdot}, in which we overlay other neutron star populations.

{\it Swift}/{\rm XRT} data were processed using 
\verb'HEASOFT v6.24', \verb'xrtpipeline v0.13.4' 
and the Calibration Database files of version 20160731.
Standard level-2 cleaned events files were generated for further analysis.
We did not use data that suffered from stray light. 
\defcitealias{XRT}{Paper~I}
\begin{table*}
\begin{center}
\caption{Summary of the {\it Swift}/{\rm XRT} X-ray archival data used.}
\label{tab:arc}
\begin{tabular}{c|c|c|c|ccc}
\hline\hline
PSR name & Date$^a$& number of the data & XRT exposure\\
&(start-end) & & ks\\
\hline
J0534$-$6703 & 2011 Oct 20 $-$ 2015 Sep 01 & 4 & 3.23  \\ 
J1107$-$6143 & 2010 Aug 04 $-$ 2011 Jan 27 & 3 &4.36 \\
J1307$-$6318 & 2011 Feb 18 $-$ 2011 Sep 30 & 2 & 1.20 \\
J1632$-$4818 & 2012 Jun 13 $-$ 2011 Sep 30 & 1 & 0.48 \\
J1713$-$3844 & 2012 Aug 11 $-$ 2011 Sep 30 & 1 & 0.57\\
J1746$-$2850 & 2011 Aug 18 $-$ 2016 Jun 06 & 921$^{b}$ & 866$^{b}$ &\\
J1755$-$2521 & 2012 Sep 07 $-$ 2016 Jun 06 & 1 & 0.31  \\
J1822$-$1252 & 2012 Oct 25 $-$ 2012 Nov 01 & 4 &2.10 \\
J1830$-$1135 & 2012 Nov 04 $-$ 2012 Nov 01 & 1 & 0.50 \\
J1851$+$0118 & 2012 Nov 20 $-$ 2012 Feb 20 & 3 & 1.15\\
J1854$+$0306 & 2013 Mar 25 $-$ 2012 Feb 20 & 1 &0.51 \\
J1855$+$0527 & 2011 Nov 16 $-$ 2013 Mar 07 & 3 &1.50\\
J1858$+$0241 & 2013 Mar 27 $-$ 2013 Mar 24 & 7 &4.58\\
J1901$+$0413 & 2012 Feb 18 $-$ 2013 Mar 24 & 1 &0.46 \\
J1905$+$0616 & 2012 Nov 29 $-$ 2012 Feb 18 & 2 &1.39 \\
J1924$+$1631 & 2012 Apr 10 $-$ 2012 Apr 10 & 2& 1.08 \\
\hline\hline
\multicolumn{5}{p{38em}}{$^a $When a PSR has multiple data, the start date of the first observation and the end date of
the finial overvation are given.}\\
\multicolumn{5}{p{38em}}{$^b$ The values are for the analysis of the 0.3$-$1.5~keV and the 0.3$-$3.0~keV bands, but for the analysis of the 0.3$-$10~keV band, 
the number of data is 920 and the total exposure time is 865~ks.}\\
\end{tabular}
\end{center}
\end{table*}

\defcitealias{XRT}{Paper~I}
\begin{table*}
\caption{Summary of our X-ray observations with {\it Swift}/{\rm XRT}. }
\label{tab:obs}
\begin{tabular}{c|c|c|c|c|c|c}
\hline\hline
PSR name & Date & ObsID & XRT exposure & off-axis\\
& (start-end) & & (ks) & (arcmin) \\
\hline
B052$5+$21 & 2016 Aug 20 $-$ 2016 Aug 20 & 00034683001 & 1.80 & 1.34\\
B1727$-$47 & 2016 May 22 $-$ 2016 May 23 & 00034538001 & 1.77 & 0.60\\
J0140$+$5622$^a$  & 2016 Jun 03 $-$ 2016 Jun 04 & 00034537001 & 2.89 & 1.82\\
J1558$-$5756 & 2016 May 11 $-$ 2016 May 11 & 00034536001 & 0.07 & 4.58\\
& 2016 May 21 $-$ 2016 May 21 & 00034536002 & 0.46& 1.44\\
& 2016 Jun 16 $-$ 2016 Jun 16 & 00034536003 & 1.20 & 0.95\\
J1840$-$0840 & 2016 May 21 $-$ 2016 May 21 & 00034540001 & 0.27& 0.76\\
& 2016 Jun 13 $-$ 2016 Jun 13 & 00034540002 & 1.78 & 1.34\\
J1851$+$0118$^b$ & 2018 Feb 28 $-$ 2018 Feb 28 & 00010585001 & 5.06 & 2.53\\
\hline\hline
\multicolumn{6}{p{38em}}{$^a$ J0140$+$5622 is cited in the ATNF pulsar catalogue as J0139$+$5621.}\\
\multicolumn{6}{p{38em}}{$^b$ We performed a 2nd observation of PSR J1851$+$0118 because it gave a 4$\sigma$ detection in the first analysis of the archival data.}\\
\end{tabular}
\end{table*}
\section{Analysis and Results}
For the X-ray counterpart search, significance against the background 
depends on the choice of energy band. 
We suppose three types of spectral models in order to optimise the energy 
range for count extraction.
The first type is a low-temperature high-B PSR model, which is a single 
blackbody model with $kT=0.15$~keV. 
The second is a high-temperature magnetar-like high-B PSR model, 
which is a single blackbody model with $kT=0.3$~keV. 
The final type is an ordinary PSR model, 
which is a single power law model with a photon index of $1.5$. 
The assumed temperature in the first model is the average value of 
the high-B PSRs 
PSR~J0726$+$2612 \citep{Speagle2011}, 
PSR~J1819$-$1458 \citep{McLaughlin2007}, 
PSR~J1718$-$3718 \citep{Zhu2011}, and 
PSR~J1119$-$6127 \citep{2012ApJ...761...65N}.
The temperature in the second model is that of PSR~J1734$-$3333 \citep{Olausen2013}, which has the highest temperature among detected high-B PSRs.
This value also sometimes appears in persistent magnetars.
For the photon index in the third model, we used the average of the high-B PSRs 
PSR~J1930$+$1852 \citep{2007ApJ...663..315L}, 
PSR~J1124$-$5916 \citep{2001ApJ...559L.153H},
PSR~B1509$-$58 \citep{KP08}, and
PSR~J1119$-$6127 \citep{2012ApJ...761...65N}.

Given a fake spectrum for each of the three types of objects and a
background spectrum, we calculated S/N ratio as a function of the energy bands. 
As a result, the
optimized energy bands were determined to be 
the $0.3-1.5$~keV band for the low-temperature high-B PSR model, 
the $0.3-3.0$~keV band for the high-temperature magnetar-like high-B 
PSR model, 
and the $0.3-10$~keV band for the ordinary PSRs model.
The source region was assumed to be circular with a 20-pixel radius (47.1 arcsec), which corresponds to the 90\% encircled energy radius at 1.5 keV. 
We obtained the background spectrum from six source-free circular regions of 20-pixels in radius, where we took data from an 8.7 ks observation of XINS RXJ0720.4$-$3125 (ObsId 00050200005), 26~ks observation of the same object (ObsId 00050203001) and the 13~ks observation of XINS RXJ1856.5$-$3754 (ObsId 00051950025).
We changed the normalization as $1\times10^{-4}$, $1\times10^{-5}$ and $1\times10^{-6}$ for the blackbody model with $kT=0.15$~keV, and the optimal energy bands were hardly changed.

In this study, 
we analysed data for these three energy bands.
We extracted source counts from each dataset using a circular region of
 20 pixels in radius centred on the radio position.
The background region was not uniquely determined for every observation, 
because the observations were not targeted to the pulsar and there were 
some number of bad pixels.
 The background region is a set of 6 circular regions with radii of 20 pixels, 
selected from 24 candidate regions to minimize bad pixels and defects 
in the field of view with 
``SAOImage DS9'' and ``Funtools''\footnote{http://hea-www.harvard.edu/saord/funtools/help.html}.

Using the correction factor yielded by \verb'xrtmkarf' for the exposure, 
vignetting, and bad columns and pixels,
we obtained corrected source and background counts.
When an object had multiple observations, 
we summed the corrected source and background counts, 
and calculated the total exposure for each object.

We evaluated the significance of source counts for every 
three bands according to Poisson statistics \citep{1986ApJ...303..336G}.
We found no detections with significance larger than $3\sigma$.
We then calculated the $3\sigma$ upper limits using Poisson 
statistics \citep{1986ApJ...303..336G}.
The obtained upper-limit count rates are given in Table~\ref{result}.
Unabsorbed flux in the $0.3-10$~keV band were also found using 
the \verb'PIMMS' (Portable Interactive Multi-Mission Simulator) 
tool\footnote{version 4.8a, http://heasarc.nasa.gov/Tools/w3pimms.html}, 
where the spectral model is the corresponding one and 
the hydrogen column density ${\it N_{\rm H}}$ is calculated 
from the dispersion measure by the empirical relation in \citet{He2013}. 
We calculated the intrinsic X-ray luminosities 
in Table~\ref{result} using distances taken from the ATNF catalogue.

\begin{figure}
\begin{center}
\includegraphics[width=0.95\columnwidth]{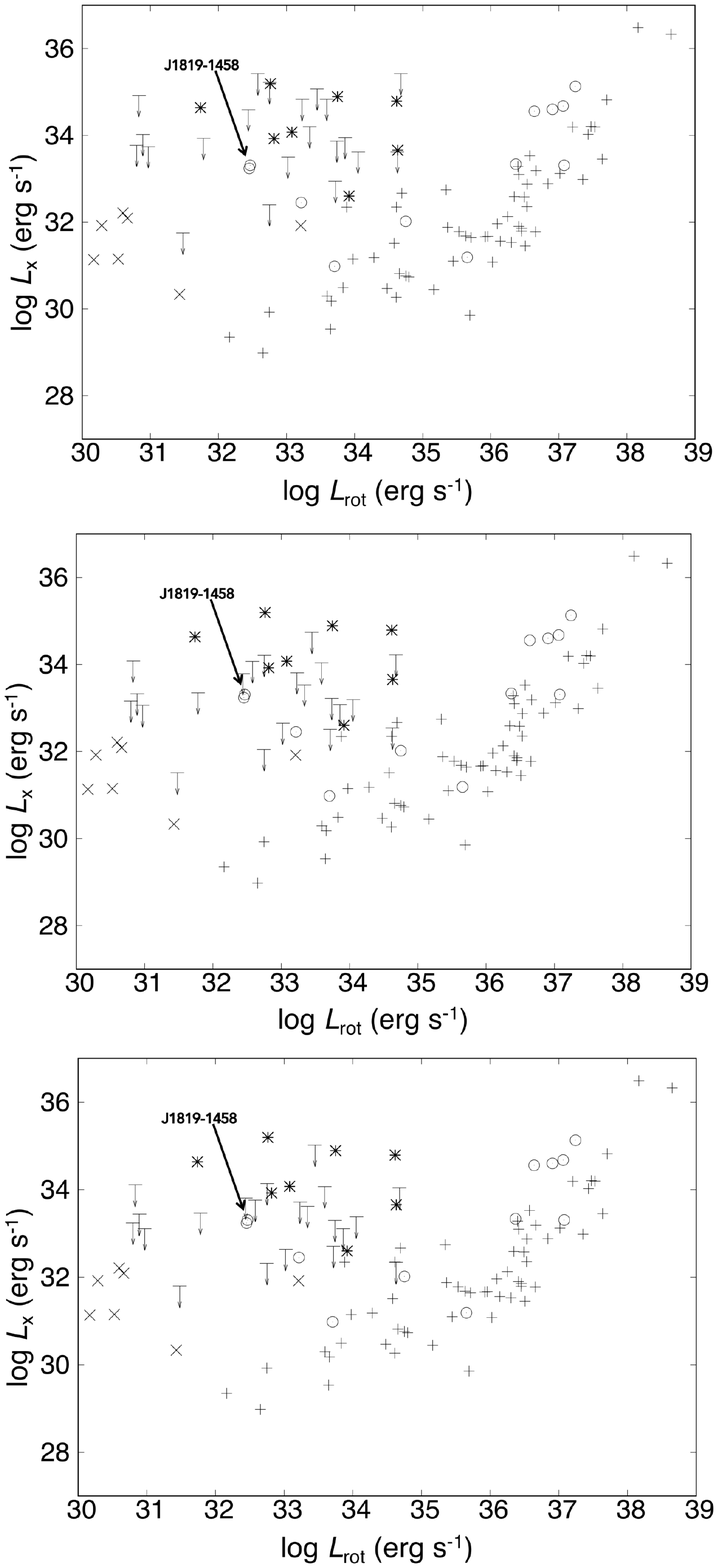}
\caption{The 3$\sigma$ upper limits are shown as arrows 
in the ${\rm log}~{\it L_{\rm x}}-{\rm log}~{\it L_{\rm rot}}$ diagram 
for the present objects with other populations in the $0.3-10$~keV band.
White circles indicate high-B PSRs detected in previous studies.
The symbol "$+$" indicates PSRs for which data were taken 
from Table~\ref{PlotData}.
Crosses and asterisks indicate XINSs \citep{vigano2013} and 
magnetars \citep{2014ApJS..212....6O} (see Table~\ref{PlotData}), respectively. 
The top, middle, and bottom panels show results 
from the low-temperature high-B PSR model, 
the high-temperature magnetar-like high-B PSR model, 
and the ordinary PSRs model, respectively. }
\label{LxLrot}
\end{center}
\end{figure}

We estimated systematic uncertainty caused by the selected 
background using unbiased variances of the background counts.
As a result, average systematic uncertainties at the $1\sigma$ 
confidence level for each band were 7.6\%, 8.0\%, and 11.7\% of 
the $3\sigma$ confidence limit.
The upper limits in terms of the X-ray luminosity ${\it L_{\rm x}}$ 
are plotted against the spin-down luminosity in Figure~\ref{LxLrot}.
One can see that some of the upper limits are less than the luminosity 
of the persistent emission of the magnetars and PSR~J1819$-$1458 having 
the highest X-ray efficiency in the high-B PSRs.

\begin{figure}
\begin{center}
\includegraphics[width=0.8\columnwidth]{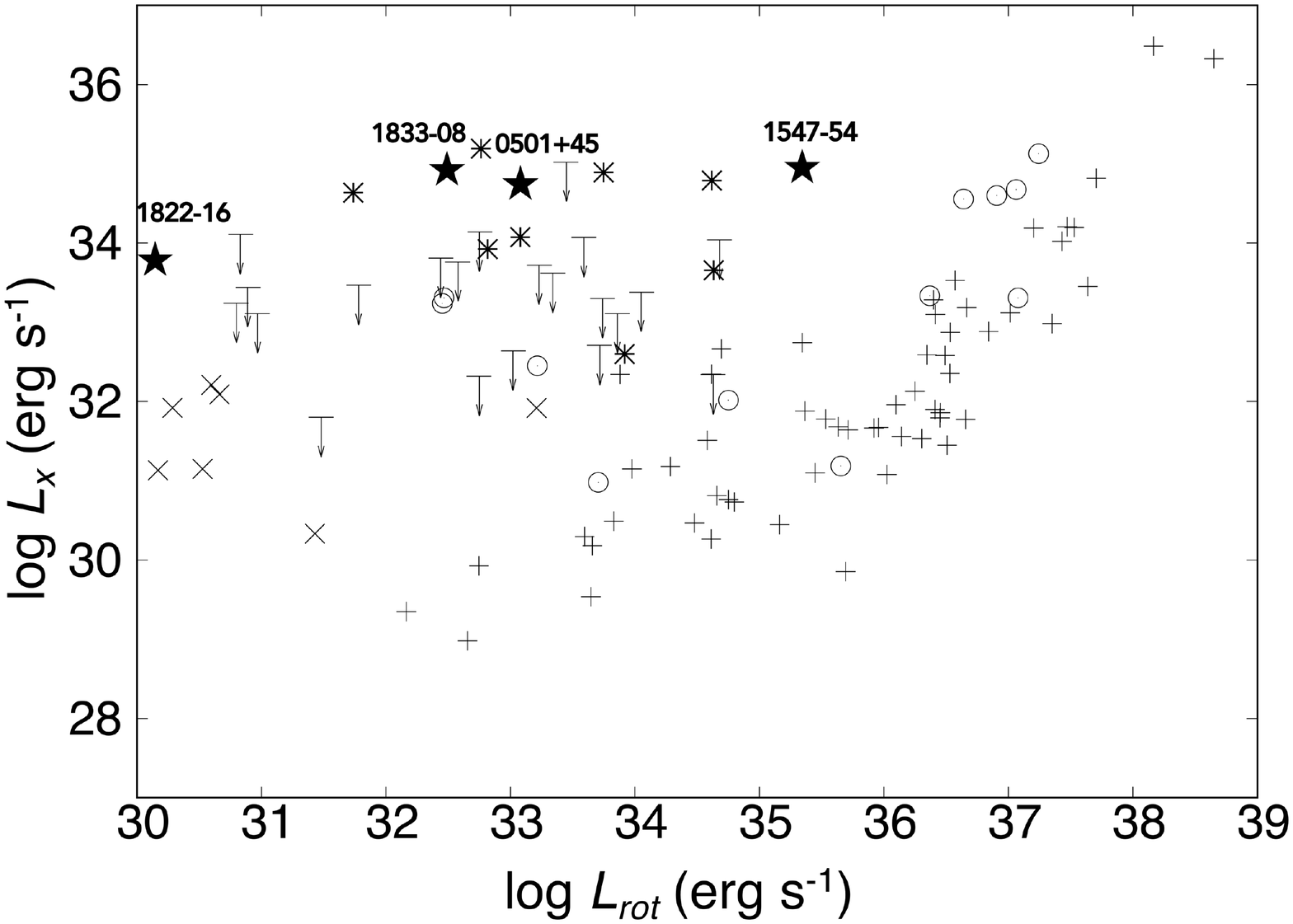}
\caption{${\rm Log}~{\it L_{\rm x}}-{\rm log}~{\it L_{\rm rot}}$ diagram 
of the power law model in the $0.3-10$~keV band for transient magnetars.
Star symbols indicate the soft-component X-ray luminosity at burst phase 
in the $1-10$~keV band of transient magnetars given in Table~6 of 
\citet{2017ApJS..231....8E}.
Other symbols are the same as in Figure~\ref{LxLrot}.
Our upper limits are smaller than the burst-phase luminosity of 
Swift~J1822.3$-$1606, except for seven upper limits.}
\label{sgr}
\end{center}
\end{figure}

In Figure~\ref{sgr},
we compare the upper limits of the ordinary PSR model with the $1.0-10$~keV 
X-ray luminosity at the burst phase of the transient magnetars 
given in Table~6 of \citet{2017ApJS..231....8E}. 
Our upper limits are smaller than the burst phase luminosity of 
Swift J1822.3$-$1606, except for seven upper limits.
Therefore, we can see that they did not magnetically explode during 
{\it Swift} observations.
\clearpage
\defcitealias{XRT}{Paper~I}
\begin{landscape}
 \begin{table}
  \caption{Summary of X-ray survey of RPPs by  {\it Swift}/{\rm XRT}}
  \label{result}
\small 
\begin{tabular}{ccccccccccccccccc}
\hline\hline
 & &&  & &  &  &&\multicolumn{3}{c}{Result of the 0.3-1.5~keV band} & \multicolumn{3}{c}{Result of the 0.3-3.0~keV band} & \multicolumn{3}{c}{Result of the 0.3-10~keV band} \\
PSR name & Exp.$^{a}$ & ${\rm log}$~${\it B_{\rm d}}$  & ${\rm log}$~${\it P}$ & ${\rm log}$~${\it P_{\rm dot}}$ & Dist. & ${\it N_{\rm H}}$ & log ${\it L_{\rm rot}}$  & cps & $F^{\rm unabs}_{\rm x}$ & ${\rm log}$~${\it L^{\rm unabs}_{\rm x}}$& cps & $F^{\rm unabs}_{\rm x}$ & ${\rm log}$~${\it L^{\rm unabs}_{\rm x}}$& cps & $F^{\rm unabs}_{\rm x}$ & ${\rm log}$~${\it L^{\rm unabs}_{\rm x}}$\\
& ks & G & s & s/s & kpc & pc ${\rm cm^{-2}}$ & erg/s & counts/s & erg~s$^{-1}$cm$^{-2}$ & erg/s& counts/s &erg~s$^{-1}$cm$^{-2}$ & erg/s& counts/s & erg~s$^{-1}$cm$^{-2}$ & erg/s\\
\hline
B0525+21 & 1.81  & 13.09  & 0.57  & -13.40  & 1.22  & 1.53$\times10^{21}$ & 31.48  & <5.34$\times10^{-3}$ & <3.13$\times10^{-13}$ & <31.75  & <5.34$\times10^{-3}$ & <1.81$\times10^{-13}$ & <31.51  & <5.78$\times10^{-3}$ & <3.53$\times10^{-13}$ & <31.80  \\
B1727-47 & 1.78  & 13.07  & -0.08  & -12.79  & 5.57  & 3.70$\times10^{21}$ & 34.05  & <8.11$\times10^{-3}$ & <1.12$\times10^{-12}$ & <33.62  & <8.36$\times10^{-3}$ & <4.19$\times10^{-13}$ & <33.19  & <8.58$\times10^{-3}$ & <6.55$\times10^{-13}$ & <33.38  \\
J0140+5622$^{b}$ & 2.90  & 13.08  & 0.25  & -13.10  & 2.41  & 3.06$\times10^{21}$ & 32.75  & <3.31$\times10^{-3}$ & <3.65$\times10^{-13}$ & <32.40  & <3.57$\times10^{-3}$ & <1.61$\times10^{-13}$ & <32.05  & <4.17$\times10^{-3}$ & <3.01$\times10^{-13}$ & <32.32  \\
J0534-6703 & 3.24  & 13.45  & 0.26  & -12.37  & 49.70  & 2.84$\times10^{21}$ & 33.45  & <3.94$\times10^{-3}$ & <4.02$\times10^{-13}$ & <35.07  & <4.33$\times10^{-3}$ & <1.89$\times10^{-13}$ & <34.74  & <5.07$\times10^{-3}$ & <3.59$\times10^{-13}$ & <35.02  \\
J1107-6143 & 4.38  & 13.23  & 0.26  & -12.81  & 2.94  & 1.22$\times10^{22}$ & 33.02  & <2.97$\times10^{-3}$ & <3.05$\times10^{-12}$ & <33.50  & <3.34$\times10^{-3}$ & <4.36$\times10^{-13}$ & <32.65  & <3.65$\times10^{-3}$ & <4.23$\times10^{-13}$ & <32.64  \\
J1307-6318 & 1.21  & 13.02  & 0.70  & -13.68  & 10.78  & 1.12$\times10^{22}$ & 30.83  & <6.98$\times10^{-3}$ & <5.99$\times10^{-12}$ & <34.92  & <7.30$\times10^{-3}$ & <8.73$\times10^{-13}$ & <34.08  & <8.20$\times10^{-3}$ & <9.21$\times10^{-13}$ & <34.11  \\
J1558-5756 & 1.73  & 13.17  & 0.05  & -12.73  & 3.32  & 3.83$\times10^{21}$ & 33.72  & <4.54$\times10^{-3}$ & <6.55$\times10^{-13}$ & <32.94  & <4.80$\times10^{-3}$ & <2.46$\times10^{-13}$ & <32.51  & <5.03$\times10^{-3}$ & <3.88$\times10^{-13}$ & <32.71  \\
J1632-4818 & 0.48  & 13.37  & -0.09  & -12.19  & 5.31  & 2.27$\times10^{22}$ & 34.68  &<1.47$\times10^{-2}$ & <7.91$\times10^{-11}$ & <35.42  & <1.73$\times10^{-2}$ & <4.95$\times10^{-12}$ & <34.22  & <2.13$\times10^{-2}$ & <3.26$\times10^{-12}$ & <34.04  \\
J1713-3844 & 0.57  & 13.23  & 0.20  & -12.75  & 4.43  & 1.63$\times10^{22}$ & 33.23  & <1.39$\times10^{-2}$ & <2.88$\times10^{-11}$ & <34.83  & <1.52$\times10^{-2}$ & <2.78$\times10^{-12}$ & <33.81  & <1.72$\times10^{-2}$ & <2.26$\times10^{-12}$ & <33.72  \\
J1746-2850$^{c}$ & 870.0$^{d}$  & 13.59  & 1.08  & 0.00  & 5.61  & 2.89$\times10^{22}$ & 34.63  & <9.10$\times10^{-5}$ & <1.11$\times10^{-12}$ & <33.62  & <2.25$\times10^{-4}$ & <9.29$\times10^{-14}$ & <32.54  & <3.36$\times10^{-4}$ & <5.79$\times10^{-14}$ & <32.34  \\
J1755-2521 & 0.31  & 13.02  & 0.07  & -13.04  & 3.44  & 7.56$\times10^{21}$ & 33.34  & <2.80$\times10^{-2}$ & <1.12$\times10^{-11}$ & <34.20  & <2.92$\times10^{-2}$ & <2.41$\times10^{-12}$ & <33.53  & <3.03$\times10^{-2}$ & <2.93$\times10^{-12}$ & <33.62  \\
J1822-1252 & 2.11  & 13.13  & 0.32  & -13.07  & 6.38  & 2.78$\times10^{22}$ & 32.58  & <5.15$\times10^{-3}$ & <5.41$\times10^{-11}$ & <35.42  & <6.23$\times10^{-3}$ & <2.41$\times10^{-12}$ & <34.07  & <7.09$\times10^{-3}$ & <1.20$\times10^{-12}$ & <33.76  \\
J1830-1135 & 0.50  & 13.24  & 0.79  & -13.32  & 3.76  & 7.71$\times10^{21}$ & 30.89  & <1.50$\times10^{-2}$ & <6.21$\times10^{-12}$ & <34.02  & <1.50$\times10^{-2}$ & <1.27$\times10^{-12}$ & <33.33  & <1.68$\times10^{-2}$ & <1.63$\times10^{-12}$ & <33.44  \\
J1840-0840 & 2.06  & 13.06  & 0.73  & -13.63  & 5.10  & 8.16$\times10^{21}$ & 30.80  & <4.20$\times10^{-3}$ & <1.92$\times10^{-12}$ & <33.77  & <5.30$\times10^{-3}$ & <4.69$\times10^{-13}$ & <33.16  & <5.64$\times10^{-3}$ & <5.60$\times10^{-13}$ & <33.24  \\
J1851+0118 & 4.63  & 13.05  & -0.04  & -12.86  & 5.64  & 1.25$\times10^{22}$ & 33.86  & <2.14$\times10^{-3}$ & <2.35$\times10^{-12}$ & <33.95  & <2.37$\times10^{-3}$ & <3.19$\times10^{-13}$ & <33.08  & <2.88$\times10^{-3}$ & <3.38$\times10^{-13}$ & <33.11  \\
J1854+0306 & 0.51  & 13.42  & 0.66  & -12.84  & 4.49  & 5.77$\times10^{21}$ & 31.78  & <1.39$\times10^{-2}$ & <3.55$\times10^{-12}$ & <33.93  & <1.39$\times10^{-2}$ & <9.29$\times10^{-13}$ & <33.35  & <1.39$\times10^{-2}$ & <1.22$\times10^{-12}$ & <33.47  \\
J1855+0527 & 1.51  & 13.29  & 0.14  & -12.57  & 11.70  & 1.09$\times10^{22}$ & 33.59  & <5.22$\times10^{-3}$ & <4.18$\times10^{-12}$ & <34.83  & <5.74$\times10^{-3}$ & <6.64$\times10^{-13}$ & <34.04  & <6.48$\times10^{-3}$ & <7.18$\times10^{-13}$ &<34.07  \\
J1858+0241 & 4.60  & 13.03  & 0.67  & -13.61  & 5.15  & 1.01$\times10^{22}$ & 30.97  & <2.56$\times10^{-3}$ & <1.76$\times10^{-12}$ & <33.74  & <3.35$\times10^{-3}$ & <3.61$\times10^{-13}$ & <33.06  & <3.79$\times10^{-3}$ & <4.08$\times10^{-13}$ & <33.11  \\
J1901+0413 & 0.47  & 13.28  & 0.43  & -12.88  & 5.34  & 1.06$\times10^{22}$ & 32.44  & <1.51$\times10^{-2}$ & <1.15$\times10^{-11}$ & <34.59  & <1.62$\times10^{-2}$ & <1.82$\times10^{-12}$ & <33.79  & <1.73$\times10^{-2}$ & <1.90$\times10^{-12}$ & <33.81  \\
J1905+0616 & 1.39  & 13.07  & 0.00  & -12.87  & 4.95  & 7.68$\times10^{21}$ & 33.74  & <6.23$\times10^{-3}$ & <2.56$\times10^{-12}$ & <33.87  & <6.77$\times10^{-3}$ & <5.68$\times10^{-13}$ & <33.22  & <7.03$\times10^{-3}$ & <6.82$\times10^{-13}$ & <33.30  \\
J1924+1631 & 1.08  & 13.52  & 0.47  & -12.44  & 10.19  & 1.56$\times10^{22}$ & 32.75  & <7.29$\times10^{-3}$ & <1.34$\times10^{-11}$ & <35.22  & <7.66$\times10^{-3}$ & <1.32$\times10^{-12}$ & <34.21  & <8.70$\times10^{-3}$ & <1.12$\times10^{-12}$ & <34.14  \\
\hline \hline 
\multicolumn{13}{l}{Notes. All upper-limits are $3\sigma$ upper-limits.}\\
\multicolumn{13}{l}{The X-ray flux and the X-ray luminosity is the unabsorbed X-ray flux and the intrinsic X-ray luminosity in the 0.3$-$10~keV band.}\\
\multicolumn{13}{l}{$^a$ The total exposure time after the data reduction.}\\
\multicolumn{10}{l}{$^b$ J0140+5622 is cited in the ATNF pulsar catalogue as J0139+5621.}\\
\multicolumn{10}{p{38em}}{$^c$ Result with Chandra is given by \citet{2017MNRAS.468.1486D}.}\\
\multicolumn{10}{p{38em}}{$^d$ When the 0.3-10~keV band is analysed, this value is 869.00.}\\
 \end{tabular}
 \end{table}
\end{landscape}

\section{Discussion}
In this section, we discuss properties of the high-B RPPs.
In subsection~4.1 we discuss dependence on the dipole magnetic field, 
and in subsection~4.2 
we estimate a fraction of the high-efficiency high-B PSRs 
in the PSR population.
\subsection{Dependence of dipole magnetic field on flux}
We found that magnetar-like properties appear only when ${\it B_{\rm d}}\gtrsim10^{13.5}$~G for the radio pulsar population. This is true even if the radio-quiet high-B RPP are included.
To see this, 
we compiled the detected X-ray flux of high-B RPPs and ordinary PSRs 
in the literature.
For the dataset of ordinary PSRs, 
we restrict ourselves to pulsars with ``rotational energy flux,'' defined as ${\it F_{\rm rot}}={\it L_{\rm rot}}/4\pi D^{2}$ larger than $1\times10^{-11}$~erg~s$^{-1}$~cm$^{-2}$,
because these would show detectable flux if they have 
an excess as compared with typical ordinary pulsars.
We also prepared data for magnetars and XINSs given 
in \citet{2014ApJS..212....6O} and \citet{vigano2013}.
The $0.3$-$10$~keV luminosities are given in Table~\ref{PlotData}.
We also utilise  upper-limit data satisfying the conditions 
that the distance is less than 4~kpc and the exposure time is longer than 1~ks.
There are 12 upper limits in total, 
of which 4 are our results and 8 are taken from previous reports.
We do not use data where too few spectral parameters are given 
for the band conversion.
\begin{figure}
\includegraphics[width=0.5\columnwidth]{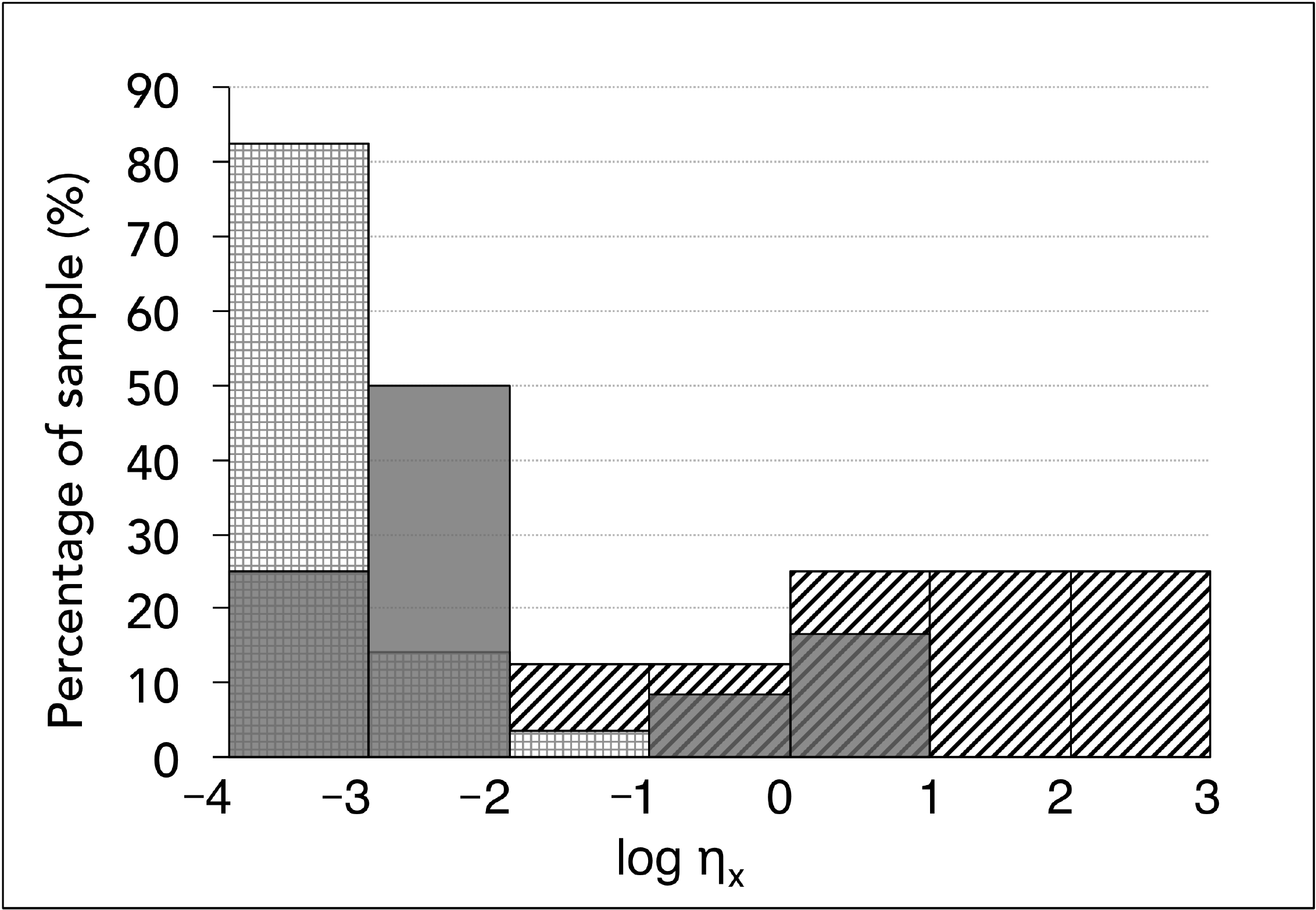}
\caption{Efficiency distributions for RPPs (checked bars), magnetars (striped bars), and high-B RPPs (dark grey bars). The data are based on Table~\ref{PlotData}.}
\label{6_histogram}
\end{figure}

\begin{figure}
\includegraphics[width=1.0\columnwidth]{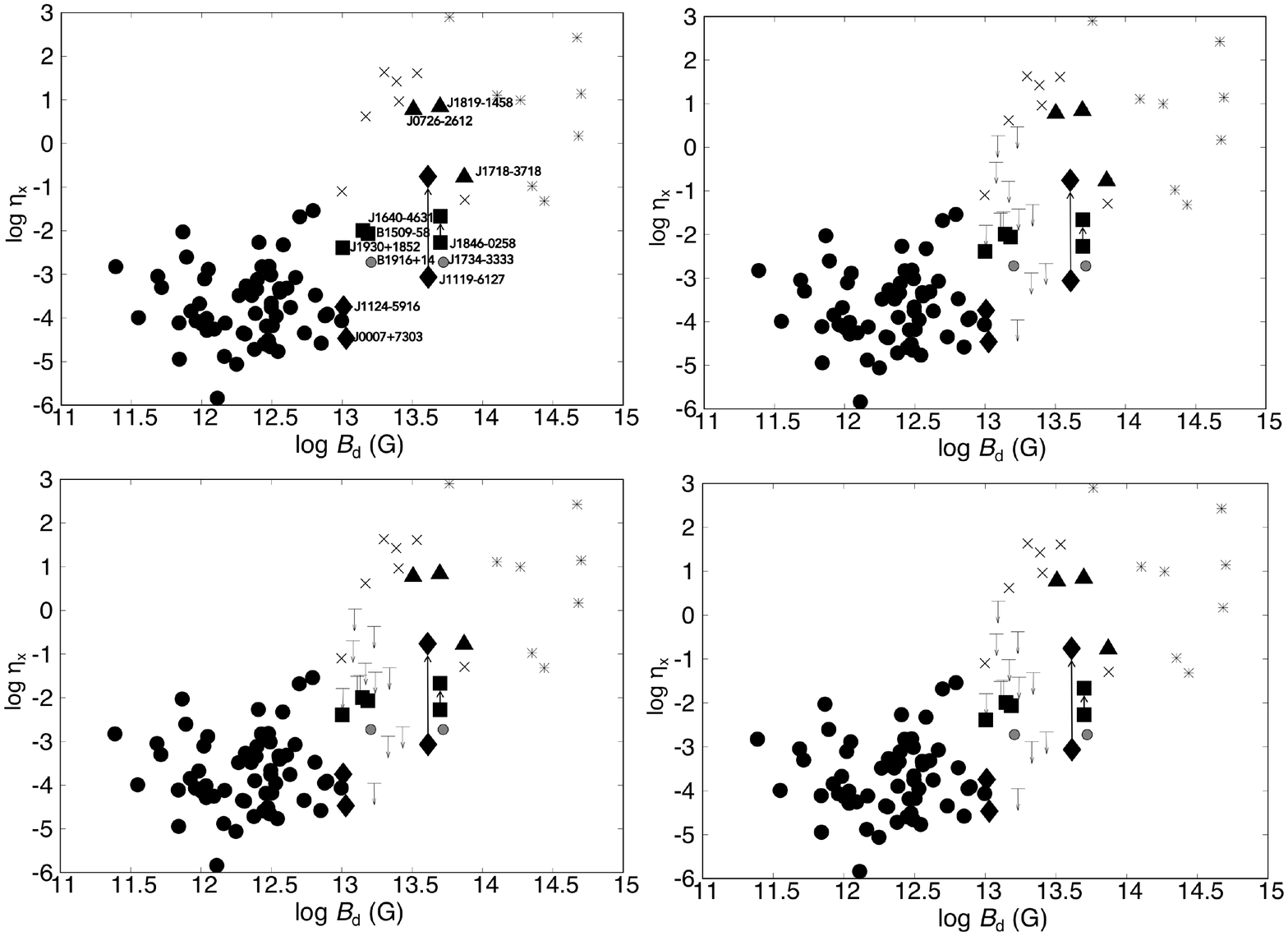}
\caption{Inferred dipole magnetic field versus X-ray efficiency.
Black filled circles indicate PSRs with ${\it B_{\rm d}}<10^{13.5}$~G, 
and triangles, rectangles, diamonds and grey circles indicate high-B RPPs, 
defined as ${\it B_{\rm d}}\geq10^{13.5}$~G, 
where the different symbols correspond to the different manifestations,
namely, triangular ones are high X-ray efficiency pulsars, rectangular ones are soft-gamma-ray pulsars, diamond ones are the gamma-ray pulsars observed with {\it Fermi}/{\rm LAT} and grey circular ones are low X-ray efficiency and a simple blackbody pulsars.
Crosses and asterisks indicate XINSs and magnetars, respectively.
Data are based on Table~\ref{PlotData}.
Top-right, bottom-left, and bottom-right panels are the same 
as the top-left panel, except that the upper limits on 4 of our results 
and 8 previous results are overlaid with arrows.
The 8 previous results are given 
in \citet{Prinz2015}, \citet{2000ApJ...528..436P}, 
and \citet{2011PASJ...63S.865U}.
Top-right, bottom-left, and bottom-right panels indicate results from 
the blackbody model with $kT= 0.15$~keV, that with $kT= 0.3$~keV, 
and the power law model with photon index $1.5$.}
\label{7_Bdefficiency}
\end{figure}
First, we plot the distribution of the X-ray efficiency 
$\eta_{x} ={\it L_{\rm x}}/{\it L_{\rm rot}}$ in the $0.3$-$10$~keV band 
for the three classes, 
high-B RPPs, ordinary PSRs, and magnetars in Figure~\ref{6_histogram}.
The efficiencies of the magnetars (striped histograms in Figure~\ref{6_histogram}) 
are more than 0.01\%, while those of the ordinary PSRs (checked) 
are less than $0.1\%$. 
The high-B RPPs (dark grey) seem to exhibit a bimodal distribution:
one is a high-efficiency group where $\eta_{x}>0.1\%$ and 
the other is a low-efficiency group where $\eta_{x}<0.01\%$.

It was suggested in \citet{2010ApJ...725..985O} that X-ray luminosity 
or efficiency weakly correlates with the dipole field ${\it B_{\rm d}}$ 
across wider neutron star populations.
We provide a ${\rm log}~\eta_{\rm x}$ versus ${\rm log}~{\it B_{\rm d}}$ plot 
in Figure~\ref{7_Bdefficiency}.
In the panels of Figure~\ref{7_Bdefficiency}, 
the distribution of 
ordinary radio pulsars (filled circles)
and 
high-B RPPs (filled triangles, rectangles, diamonds, and grey circles)
seems to show a stepwise structure 
in $\eta_{\rm x}$ at ${\it B_{\rm d}}\sim 10^{13.5}$~G. 

On the right side, 
where ${\it B_{\rm d}}>10^{13.5}$~G, 
all high-B RPPs show some features of magnetars, 
such as high efficiency or burst activity,
except for PSR~J1734$-$3333 (discussed below).
On the left side, where ${\it B_{\rm d}}<10^{13.5}$~G, 
these high-B RPPs behave like ordinary pulsars where $\eta_{\rm x}$ is low.

For the case of ${\it B_{\rm d}}>10^{13.5}$~G, PSR~J1718$-$3718, 
PSR~J0726$-$2612, and PSR~J1819$-$1458 show high X-ray efficiency 
at the relatively high temperature of $0.08$-$0.18$~keV. 
The X-ray pulse fractions of PSR~J1718$-$3718 and PSR~J1819$-$1458 are 
$52\pm13\%$ in the $0.8$-$2.0$~keV band and $34\pm6\%$ in the $0.3$-$5$~keV 
band \citep{Zhu2011, McLaughlin2007}.
As for PSR~J0726$-$2612, 
the semi-amplitude is given to be $27\pm5\%$ in the $0.32$-$1.1$~keV 
band \citep{Speagle2011}.
These properties are an indication of magnetic activity on the surface.
Burst activity was found in PSR~J1846$-$0258 and PSR~J1119$-$6127.
The former shows a magnetar-like outburst in 2006. 
In quiescence, it is a soft-gamma-ray pulsar with pulsar wind nebula.
The efficiency is ${\rm log}~\eta_{\rm x}=-2.52$ in the quiescent state 
and becomes ${\rm log}~\eta_{\rm x}=-1.73$ at the burst state 
\citep{2008ApJ...686..508N}.
The latter also shows a magnetar-like outburst in 2016, 
and observations by {\it Fermi}/{\rm LAT} suggest 
it to be an ordinary RPP.
The efficiency is ${\rm log}~\eta_{\rm x}=-3.11$ in 
quiescence and became ${\rm log}~\eta_{\rm x}=-0.82$ during 
the burst \citep{2012ApJ...761...65N, Archibald2016}.

Members in the low-field side, ${\it B_{\rm d}}<10^{13.5}$~G, 
manifest themselves differently. 
(1) PSR~B1509$-$58, PSR~J1930$+$1852, and PSR~J1640$-$4631 are 
referred to as soft-gamma-ray pulsars, 
indicated by rectangles in Figure~\ref{7_Bdefficiency} \citep{kuiper2015}, 
(2) PSR~J0007$+$7303 and PSR~J1124$-$5916 are classified in the gamma-ray 
pulsars observed with {\it Fermi}/{\rm LAT}, indicated by 
diamonds \citep{2013ApJS..208...17A}, and 
(3) PSR~B1916$+$14, indicated by the filled grey circle, 
shows low efficiency and a simple blackbody spectrum 
\citep{Olausen2013, Zhu2009}.

The meaning of these apparent differences is not clear.
However, they would not be related to the strength of the dipole fields, 
because 
(1)~the dipole field strength of the soft gamma-ray pulsars are 
widely distributed, from $7.5\times10^{11}$~G to $4.9\times10^{13}$~G,
(2)~{\it Fermi}/{\rm LAT} pulsars are very common in RPPs with various 
field strengths, and
(3)~thermally dominant X-ray radiation is also common in middle-aged RPPs.

In summary, the three highly efficient and two bursting high-B RPPs 
all have larger dipole fields than a threshold 
at ${\it B_{\rm d}}\sim10^{13.5}$~G.
This argument may be strengthened if we add the high-B RPPs 
that have tight upper limits.
We plotted the upper limits with arrows in the top-right, 
bottom-left, and bottom-right panels in Figure~\ref{7_Bdefficiency}.
Some upper limits confirm that objects are definitely not 
as X-ray efficient as PSR~J1718$-$3718 and rather like a normal PSR.
There are 9 such objects.
\begin{figure}
\includegraphics[width=0.8\columnwidth]{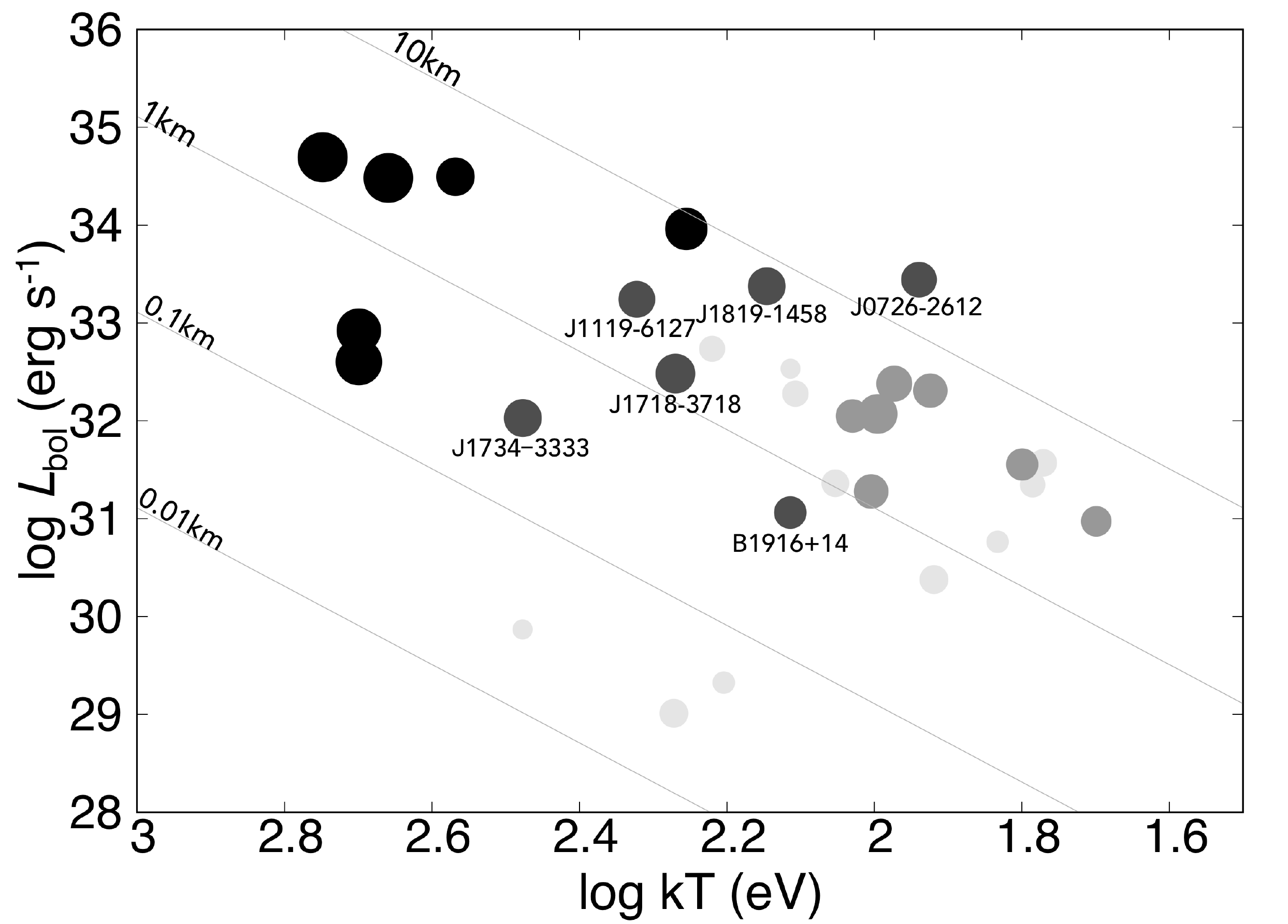}
\caption{${\rm Log}~{\it kT}$-${\rm log}~{\it L_{\rm bol}}$ diagram, 
where $kT$ and ${\it L_{\rm bol}}$ are the blackbody temperature and 
the bolometric luminosity from Table~\ref{PlotDataBB} and Table~\ref{PlotData}.
The light grey, grey, dark grey, and black circles indicate PSRs, XINSs, high-B RPPs, and magnetars, respectively.
Circle sizes indicate strength of the dipole magnetic field.
Lines show the radiative radius calculated as $\sqrt{{{\it L_{\rm x}}\slash 4\pi \sigma T^{4}}}$.}
\label{9_BBLxkT}
\end{figure}
\defcitealias{XRT}{Paper~I}
\begin{table*}
\caption{PSRs and high-B RPPs with blackbody components.}
\label{PlotDataBB}
\small
\begin{tabular}{c c c c c c c c c c c c}
\hline\hline
PSR name & type &${\it B_{\rm d}}$&  Dist. & Temp. &  model$^{a}$ & component & log${\it L_{\rm bol}}$ & log${\it L_{\rm rot}}$ & Reference\\
&&G& kpc & keV  & &  &erg/s & erg/s &\\
\hline
 B0656+14 &B0 &4.66e+12 & 0.29 & 0.059 &  PL+BB+BB &Blackbody 1 & 31.57 & 34.58 &  \citet{2016ApJ...817..129B} \\
 B0656+14 &B0 &4.66e+12 &  0.29 & 0.113 & PL+BB+BB &Blackbody 2 & 31.36 & 34.58 & \citet{2016ApJ...817..129B}  \\
 B0833-45 & B0&3.39e+12 &  0.28 & 0.128 & PL+BB & Blackbody  &32.28 & 36.84 &  \citet{Pavlov2001} \\
 B1055-52 & B0&1.09e+12 & 0.09 & 0.068 &  PL+BB+BB &Blackbody 1 & 30.76 & 34.48 &  \citet{2015ApJ...811...96P}   \\
 B1055-52 & B0&1.09e+12 & 0.09 & 0.16 & PL+BB+BB &Blackbody 2 & 29.33 & 34.48 & \citet{2015ApJ...811...96P}   \\
 B1706-44 & B0&3.13e+12 &  2.60 & 0.166 & PL+BB &Blackbody &32.74 & 36.53 &  \citet{2005ApJ...631..480R}  \\
 B1822-09 & B0&6.44e+12 &  0.30 & 0.083 &  BB+BB &Blackbody 1 & 30.38 & 33.66 &  \citet{Hermsen2017}\\
 B1822-09 & B0&6.44e+12 &  0.30 & 0.187 & BB+BB &Blackbody 2 & 29.01 & 33.66 & \citet{Hermsen2017} \\
 B1916+14 & B0&1.60e+13 &1.30 & 0.13 &  BB &Blackbody & 31.06 & 33.71 &  \citet{Zhu2009}  \\
 B1929+10 & B0&5.19e+11 &0.31 & 0.30 & PL+BB & Blackbody & 29.87 & 33.60 &  \citet{2008ApJ...685.1129M} \\
 B1951+32 & B0&4.86e+11 & 3.00 & 0.13 & PL+BB &Blackbody &  32.53 & 36.57 &  \citet{2005ApJ...628..931L}  \\
 J1741-2054 & B0&2.69e+12 &  0.30 & 0.061 &  PL+BB &Blackbody & 31.35 & 33.98 &\citet{2014ApJ...790...51M} \\
\hline 
J0726-2612 & HB&3.21e+13 & 2.90 & 0.087 &  BB &Blackbody & 33.44 & 32.45 &  \citet{Speagle2011}  \\
 J1119-6127 & HB&4.10e+13 &  8.40 & 0.21 & PL+BB &Blackbody& 33.24 & 36.37 &  \citet{2012ApJ...761...65N} \\
 J1718-3718 & HB&7.46e+13 &  3.92 & 0.186 &  BB &Blackbody & 32.48 & 33.22 & \citet{Zhu2011} \\
 J1734-3333 & HB&5.24e+13 & 4.46 & 0.30 &  BB &Blackbody & 32.03 & 34.75 & \citet{Olausen2013}  \\
 J1819-1458 & HB&5.01e+13 &  3.30 & 0.14 &  BB & Blackbody & 33.38 & 32.47 &  \citet{McLaughlin2007}\\
 \hline 
J0501+4516	&	Mag	&	1.85E+14	&	2.20	&	0.5	&	PL+BB	&	Blackbody &	32.92&33.08	&\citet{2014MNRAS.438.3291C} \\
J1050-5953	&	Mag	&	5.02E+14	&	9.00	&	0.56	&	PL+BB	&	Blackbody &	34.69	&	33.75	&\citet{2008ApJ...677..503T} \\
 J1622-4950	&	Mag	&	2.75E+14	&	5.57	&	0.5	&	BB	&	Blackbody &	32.60	&	33.92	&\citet{2012ApJ...751...53A} \\
J1708-4008	&	Mag	&	4.70E+14	&	3.80	&	0.456	&	PL+BB	&	Blackbody &	34.48	&	32.76	&\citet{2007ApandSS.308..505R} \\
J1809-1943	&	Mag	&	1.27E+14	&	3.60	&	0.28	&	BB	&	Blackbody &	33.96	&	32.82	&\citet{2004ApJ...605..368G} \\
J2301+5852	&	Mag	&	5.81E+13	&	3.30	&	0.37	&	PL+BB	&	Blackbody &	34.49	&	31.74	&\citet{2008ApJ...686..520Z} \\
\hline\hline
\multicolumn{9}{l}{$^a$PL and BB indicate the power law model and the blackbody model. }\\
\end{tabular}
\end{table*}
Differences in populations among PSRs, high-B RPPs, 
and magnetars are clearer when we draw an ``HR diagram'' 
for neutron stars, plotting the bolometric X-ray luminosity 
versus the blackbody temperature as seen in Figure~\ref{9_BBLxkT}.
Note that one pulsar can have multiple blackbody components. 
In such cases, we plot one point for each component.
The data are given in Table~\ref{PlotDataBB}.
The points are distributed from the top left to the bottom right, forming what can be regard as a "main sequence" for pulsars. 
It is interesting to see that in this main sequence magnetars are in the top left, high-B pulsars are in the middle,
and ordinary pulsars are in the bottom right.
The points are distributed in the order of magnetic field strength, 
indicated by radii.
In the discussion of Figure~\ref{7_Bdefficiency}, 
we note that PSR~J1734$-$3333 is neither a high-efficiency object nor a burster,
 but it has a larger dipole field than the threshold.
In the HR diagram, however, PSR~J1734$-$3333 is located near the border of 
the magnetar region, meaning it has an exceptionally high temperature, so PSR~J1734$-$3333 has an indication of magnetic heating.
Therefore, PSR~J1734$-$3333 is an interesting object for future observations.
\begin{figure}
\includegraphics[width=0.8\columnwidth]{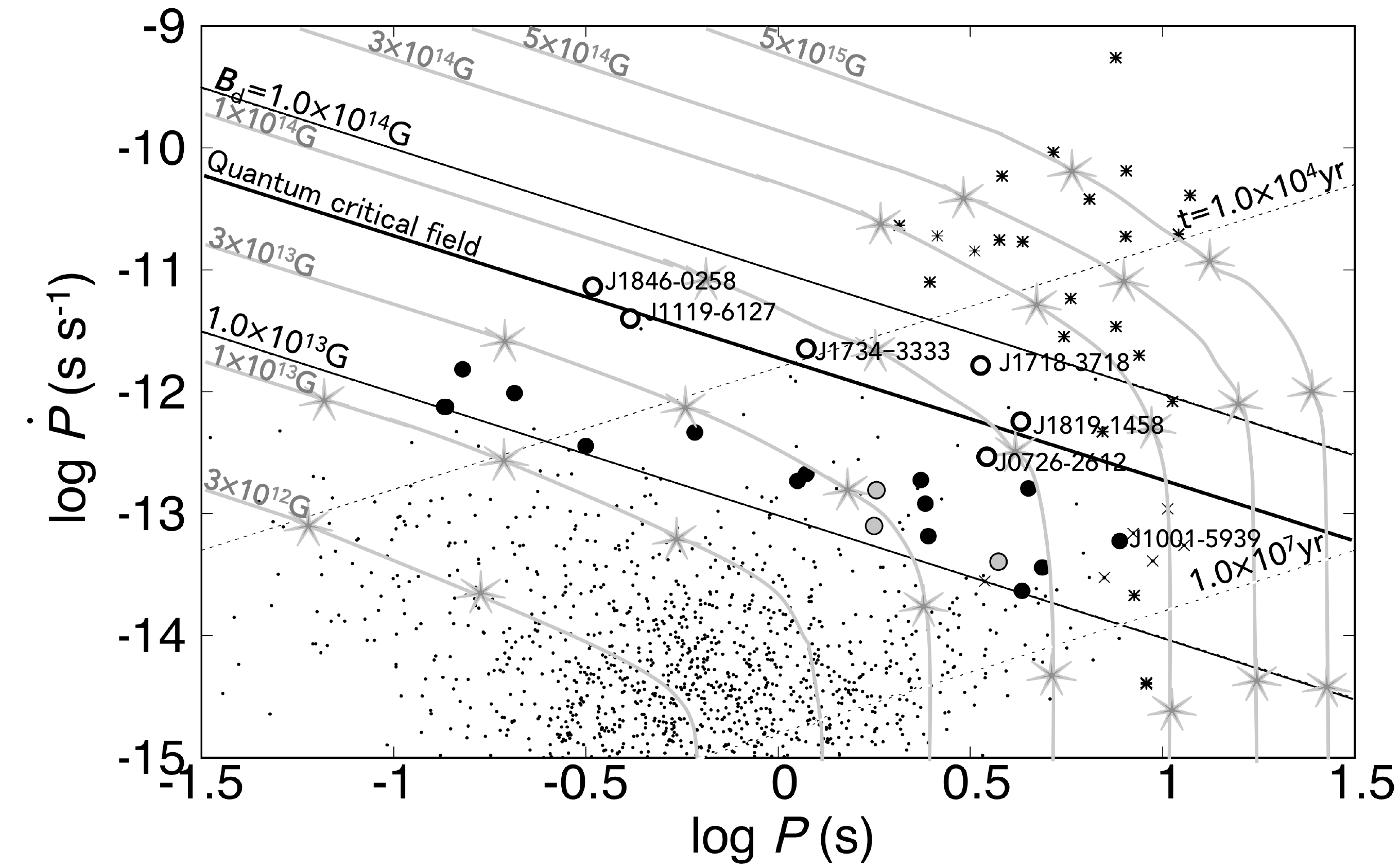}
\caption{$P-\dot{P}$ diagram for high-B RPPs with the evolutionary track given in \citet{vigano2013}.
High-B RPPs are indicated by circles, 
discriminating between objects showing magnetic activity (open circles) 
and not (filled circles). 
Grey circles are objects with weak upper limits.
Grey solid line curves indicate simulated evolutionary tracks 
labelled with the initial dipole magnetic fields ${\it B_{\rm p}^{0}}=3\times10^{12}, 10^{13}, 3\time10^{13}, 10^{14}, 3\times10^{14}, 10^{15}$G. 
Star symbols are actual ages $t = 10^{3}, 10^{4}, 10^{5}, 5\times10^{5}$yr. 
Other symbols are the same as in Figure~\ref{PPdot}.}
\label{8_PPdot-discussion}
\end{figure}

In view of theory, we draw the evolutionary tracks of different initial 
magnetic fields given in \citet{vigano2013}, 
indicated by the grey lines in the $P-\dot{P}$ diagram 
in Figure~\ref{8_PPdot-discussion}.
To see properties of high-B PSRs, 
open circles indicate high-B RPPs with magnetar-like properties such as outbursts, 
high X-ray efficiency, or high temperature, 
while black circles indicate high-B RPPs that do not show high efficiency.
Other symbols are the same as in Figure~\ref{PPdot}.
According to \citet{vigano2013}, 
RPPs having high initial magnetic fields larger than ${B_{\rm d}}\gtrsim 1 \times10^{14}$~G will be 
much more luminous 
as compared with standard cooling 
due to additional heating, 
while RPPs having initial magnetic fields lower 
than ${B_{\rm d}}\lesssim 1 \times10^{14}$~G 
will have little effect on X-ray luminosity.
Their argument seems to work almost as well.
However, 
PSR~J1001$-$5939 in the XINS domain does not follow their argument.
Some high-B RPPs with a weaker magnetic field than the quantum magnetic field 
may evolve with no significant decay of ${B_{\rm d}}$ 
along straight lines with constant magnetic fields, 
rather than their evolutional tracks.
Future observations of very high-B RPPs 
such as PSR~J1847$-$0130 and PSR~J1814$-$1744 
are of great importance toward understanding the role of 
the dipole magnetic field.

In the discussions so far, 
we have argued that magnetar-like properties appear 
only when ${\it B_{\rm d}}\gtrsim10^{13.5}$~G 
as far as high-B RPPs and detected ordinary PSRs are concerned.
However, the number of samples remains small, 
so this argument would not be conclusive. 
We should thus perform further observations in the future.
It is interesting that magnetars showing persistent emissions are found 
in the range ${\it B_{\rm d}}\gtrsim10^{13.5}$~G,
while the dipole field strengths of the XINSs extend below $10^{13.5}$~G.

\subsection{Probability of high-efficiency RPPs}
Let us calculate the probability at which high-efficiency RPPs appear among high-B RPPs.
There are 7 high-B RPPs within 2~kpc from the earth in the ATNF pulsar catalogue.
Thus, the surface density is 1.75 high-B PSRs~kpc$^{-2}$, 
provided that the radio survey is complete within 2~kpc.
Three high-efficiency high-B PSRs are located at $\sim 4$~kpc.
Using the surface density, 
the expected number of high-B PSRs within 4~kpc is calculated to be 28.
Then the probability of high efficiency appearing in the high-B PSRs 
is estimated to be 3/28, or approximately 11\%.
There are five objects having no effective upper limits within 4~kpc.
If these objects have high efficiency, 
the probability will be a higher value: 
adding these five objects, 
we would have a maximum probability of 8/28, or approximately 29\%.

\section{Conclusion}
In this work, 
we performed a systematic analysis of {\it Swift}/{\rm XRT} data 
for 21 high-B PSRs that had no previously given X-ray flux.
As a result, we newly presented $3\sigma$ upper limits for those 21 objects.
We found no sources with luminosity comparable to PSR~J1819$-$1458, 
which is a typical high-efficiency X-ray RPP. 

The present data suggest that magnetar-like properties 
appear only when ${\it B_{\rm d}}\gtrsim10^{13.5}$~G for the high-B RPPs and ordinary PSRs.
This observation is in agreement with theoretical predictions for magnetic field evolution,
namely that an evolutionary track with an initial toroidal field of 
$\sim 10^{14}$~G separates observational features 
(discussed in subsection~4.1; \citealt{vigano2013}).
It is, however, notable that a few high-B pulsars in the XINS domain 
might evolve without significant field decay, 
because of a different evolutionary history for internal magnetic fields.

An increased upper limit allowed us to estimate 
the probability of high X-ray efficiency to be 11$-$29\% in the high-B PSR population.

\section*{Acknowledgments}
This paper is based on observations obtained with {\it Swift}. 
We acknowledge the use of public data from the Swift data archive.
The authors thank the referee for his invaluable comments that improved this paper. 
We also thank H. Ohno for useful discussions of this work.
This research made use of SAOImage DS9, developed by SAO.
This study was supported in part by Grants-in-aid for Scientific Research (SS 25400221, 18H01246 and AB 15K05107) from MEXT.




\clearpage
\onecolumn
\appendix 
\section{Previous observation data for PSRs and high-B PSRs }
\label{sec:abbreviations}
Table~\ref{PlotData} provides the data used to draw Figure~\ref{LxLrot}, Figure~\ref{6_histogram}, Figure~\ref{7_Bdefficiency} and Figure~\ref{9_BBLxkT} according to our literature search.
\defcitealias{XRT}{Paper~I}
\small
\renewcommand{\arraystretch}{0.8}
\begin{longtable}{c c c c c c c c c} 
\caption{Plot data in the $0.3-10$~keV band}
\label{PlotData}\\
\hline\hline
PSR name & ${\it B_{\rm d}}$ &  ${\it P}$ & ${\it P_{\rm dot}}$ & Distance & log ${\it L_{\rm rot}}$ &  log ${\it L_{\rm x}}$ &  Type$^{a}$  & Reference\\
  & G &  s & s/s & kpc & erg/s &  erg/s && \\
\hline  \endfirsthead
\caption{(continued)}\\
\hline\hline
PSR name & ${\it B_{\rm d}}$ &  ${\it P}$ & ${\it P_{\rm dot}}$ & Distance & log ${\it L_{\rm rot}}$ &  log ${\it L_{\rm x}}$ &  Type$^{a}$  & Reference\\
  & G &  s & s/s & kpc & erg/s &  erg/s && \\
\hline \endhead
\hline \endfoot
\endlastfoot
B0114$+$58	&	7.80$\times 10^{11}$	&	1.01$\times 10^{-1}$	&	5.85$\times 10^{-15}$	&	1.770 	&	35.34 	&	32.74 	&	B0	&	\citet{Prinz2015}\\
B0355$+$54	&	8.39$\times 10^{11}$	&	1.56$\times 10^{-1}$	&	4.40$\times 10^{-15}$	&	1.000 	&	34.66 	&	30.81 	&	B0	&	\citet{2016ApJ...833..253K}\\
B0531$+$21	&	3.80$\times 10^{12}$	&	3.34$\times 10^{-2}$	&	4.21$\times 10^{-13}$	&	2.000 	&	38.65 	&	36.33 	&	B0	&	\citet{KP08}\\
B0540$+$23	&	1.97$\times 10^{12}$	&	2.46$\times 10^{-1}$	&	1.54$\times 10^{-14}$	&	1.560 	&	34.61 	&	30.26 	&	B0	&	\citet{Prinz2015}\\
B0540$-$69	&	4.99$\times 10^{12}$	&	5.06$\times 10^{-2}$	&	4.79$\times 10^{-13}$	&	49.700 	&	38.16 	&	36.49 	&	B0	&	\citet{Karret2001}\\
B0628$-$28	&	3.01$\times 10^{12}$	&	1.24	&	7.11$\times 10^{-15}$	&	0.320 	&	32.16 	&	29.35 	&	B0	&	\citet{Tepedelenl2005}\\
B0656$+$14	&	4.66$\times 10^{12}$	&	3.85$\times 10^{-1}$	&	5.50$\times 10^{-14}$	&	0.290 	&	34.58 	&	31.51 	&	B0	&	\citet{2016ApJ...817..129B}\\
B0823$+$26	&	9.64$\times 10^{11}$	&	5.31$\times 10^{-1}$	&	1.71$\times 10^{-15}$	&	0.320 	&	32.65 	&	28.98 	&	B0	&	\citet{Becker2004}\\
B0833$-$45	&	3.39$\times 10^{12}$	&	8.93$\times 10^{-2}$	&	1.25$\times 10^{-13}$	&	0.280 	&	36.84 	&	32.88 	&	B0	&	\citet{Pavlov2001}\\
B0905$-$51	&	6.90$\times 10^{11}$	&	2.54$\times 10^{-1}$	&	1.83$\times 10^{-15}$	&	0.340 	&	33.65 	&	29.53 	&	B0	&	\citet{Bogdanov2014}\\
B0906$-$49	&	1.29$\times 10^{12}$	&	1.07$\times 10^{-1}$	&	1.52$\times 10^{-14}$	&	1.000 	&	35.69 	&	29.85 	&	B0	&	\citet{Kargaltsev2012}\\
B0919$+$06	&	2.46$\times 10^{12}$	&	4.31$\times 10^{-1}$	&	1.37$\times 10^{-14}$	&	1.100 	&	33.83 	&	30.49 	&	B0	&	\citet{Prinz2015}\\
B0950$+$08	&	2.44$\times 10^{11}$	&	2.53$\times 10^{-1}$	&	2.30$\times 10^{-16}$	&	0.260 	&	32.75 	&	29.92 	&	B0	&	\citet{Becker2004}\\
B1046$-$58	&	3.49$\times 10^{12}$	&	1.24$\times 10^{-1}$	&	9.64$\times 10^{-14}$	&	2.900 	&	36.30 	&	31.53 	&	B0	&	\citet{KP08}\\
B1055$-$52	&	1.09$\times 10^{12}$	&	1.97$\times 10^{-1}$	&	5.83$\times 10^{-15}$	&	0.090 	&	34.48 	&	30.47 	&	B0	&	\citet{2015ApJ...811...96P}\\
B1221$-$63	&	1.05$\times 10^{12}$	&	2.16$\times 10^{-1}$	&	4.95$\times 10^{-15}$	&	4.000 	&	34.28 	&	31.18 	&	B0	&	\citet{Prinz2015}\\
B1338$-$62	&	7.08$\times 10^{12}$	&	1.93$\times 10^{-1}$	&	2.53$\times 10^{-13}$	&	12.590 	&	36.14 	&	31.56 	&	B0	&	\citet{Prinz2015}\\
B1706$-$44	&	3.13$\times 10^{12}$	&	1.02$\times 10^{-1}$	&	9.29$\times 10^{-14}$	&	2.600 	&	36.53 	&	32.87 	&	B0	&	\citet{2005ApJ...631..480R}\\
B1757$-$24	&	4.05$\times 10^{12}$	&	1.25$\times 10^{-1}$	&	1.28$\times 10^{-13}$	&	3.790 	&	36.41 	&	33.10 	&	B0	&	\citet{KP08}\\
B1800$-$21	&	4.29$\times 10^{12}$	&	1.34$\times 10^{-1}$	&	1.34$\times 10^{-13}$	&	4.400 	&	36.35 	&	32.59 	&	B0	&	\citet{2007ApJ...660.1413K}\\
B1822$-$09	&	6.44$\times 10^{12}$	&	7.69$\times 10^{-1}$	&	5.25$\times 10^{-14}$	&	0.300 	&	33.66 	&	30.18 	&	B0	&	\citet{Hermsen2017}\\
B1822$-$14	&	2.55$\times 10^{12}$	&	2.79$\times 10^{-1}$	&	2.27$\times 10^{-14}$	&	4.470 	&	34.61 	&	32.35 	&	B0	&	\citet{Bogdanov2014}\\
B1823$-$13	&	2.80$\times 10^{12}$	&	1.01$\times 10^{-1}$	&	7.53$\times 10^{-14}$	&	3.610 	&	36.45 	&	31.86 	&	B0	&	\citet{KP08}\\
B1853$+$01	&	7.55$\times 10^{12}$	&	2.67$\times 10^{-1}$	&	2.08$\times 10^{-13}$	&	3.300 	&	35.63 	&	31.68 	&	B0	&	\citet{KP08}\\
B1929$+$10	&	5.19$\times 10^{11}$	&	2.27$\times 10^{-1}$	&	1.16$\times 10^{-15}$	&	0.310 	&	33.60 	&	30.30 	&	B0	&	\citet{2008ApJ...685.1129M}\\
B1951$+$32	&	4.86$\times 10^{11}$	&	3.95$\times 10^{-2}$	&	5.83$\times 10^{-15}$	&	3.000 	&	36.57 	&	33.53 	&	B0	&	\citet{2005ApJ...628..931L}\\
B2334$+$61	&	9.91$\times 10^{12}$	&	4.95$\times 10^{-1}$	&	1.93$\times 10^{-13}$	&	0.700 	&	34.80 	&	30.73 	&	B0	&	\citet{McGowan2006}\\
J0205$+$6449	&	3.61$\times 10^{12}$	&	6.57$\times 10^{-2}$	&	1.94$\times 10^{-13}$	&	3.200 	&	37.43 	&	34.02 	&	B0	&	\citet{2010AandA...515A..34K}\\
J0538$+$2817	&	7.35$\times 10^{11}$	&	1.43$\times 10^{-1}$	&	3.67$\times 10^{-15}$	&	1.300 	&	34.69 	&	32.67 	&	B0	&	\citet{Ng2007}\\
J0729$-$1448	&	5.40$\times 10^{12}$	&	2.52$\times 10^{-1}$	&	1.13$\times 10^{-13}$	&	2.690 	&	35.45 	&	31.10 	&	B0	&	\citet{Kargaltsev2012}\\
J0855$-$4644	&	6.93$\times 10^{11}$	&	6.47$\times 10^{-2}$	&	7.26$\times 10^{-15}$	&	5.710 	&	36.02 	&	31.08 	&	B0	&	\citet{Maitra2017}\\
J1016$-$5857	&	2.99$\times 10^{12}$	&	1.07$\times 10^{-1}$	&	8.07$\times 10^{-14}$	&	3.160 	&	36.41 	&	31.90 	&	B0	&	\citet{KP08}\\
J1028$-$5819	&	1.23$\times 10^{12}$	&	9.14$\times 10^{-2}$	&	1.61$\times 10^{-14}$	&	1.420 	&	35.92 	&	31.67 	&	B0	&	\citet{2012AandA...543A.130M}\\
J1112$-$6103	&	1.45$\times 10^{12}$	&	6.50$\times 10^{-2}$	&	3.15$\times 10^{-14}$	&	4.500 	&	36.66 	&	31.78 	&	B0	&	\citet{Prinz2015}\\
J1301$-$6310	&	6.19$\times 10^{12}$	&	6.64$\times 10^{-1}$	&	5.64$\times 10^{-14}$	&	1.460 	&	33.88 	&	32.34 	&	B0	&	\citet{Prinz2015}\\
J1357$-$6429	&	7.83$\times 10^{12}$	&	1.66$\times 10^{-1}$	&	3.60$\times 10^{-13}$	&	3.100 	&	36.49 	&	32.58 	&	B0	&	\citet{2012ApJ...744...81C}\\
J1400$-$6325	&	1.12$\times 10^{12}$	&	3.12$\times 10^{-2}$	&	3.89$\times 10^{-14}$	&	7.000 	&	37.70 	&	34.82 	&	B0	&	\citet{Renaud2010}\\
J1420$-$6048	&	2.41$\times 10^{12}$	&	6.82$\times 10^{-2}$	&	8.32$\times 10^{-14}$	&	5.620 	&	37.02 	&	33.12 	&	B0	&	\citet{KP08}\\
J1509$-$5850	&	9.14$\times 10^{11}$	&	8.89$\times 10^{-2}$	&	9.16$\times 10^{-15}$	&	3.350 	&	35.71 	&	31.64 	&	B0	&	\citet{2016ApJ...828...70K}\\
J1524$-$5625	&	1.77$\times 10^{12}$	&	7.82$\times 10^{-2}$	&	3.90$\times 10^{-14}$	&	3.380 	&	36.51 	&	31.45 	&	B0	&	\citet{Kargaltsev2012}\\
J1531$-$5610	&	1.09$\times 10^{12}$	&	8.42$\times 10^{-2}$	&	1.37$\times 10^{-14}$	&	2.850 	&	35.96 	&	31.67 	&	B0	&	\citet{Kargaltsev2012}\\
J1617$-$5055	&	3.10$\times 10^{12}$	&	6.94$\times 10^{-2}$	&	1.35$\times 10^{-13}$	&	4.740 	&	37.20 	&	34.19 	&	B0	&	\citet{2009ApJ...690..891K}\\
J1702$-$4128 &	3.13$\times 10^{12}$	&	1.82$\times 10^{-1}$	&	5.24$\times 10^{-14}$	&	3.970 	&	35.53 	&	31.78 	&	B0	&	\citet{Kargaltsev2012}\\
J1718$-$3825	&	1.01$\times 10^{12}$	&	7.47$\times 10^{-2}$	&	1.32$\times 10^{-14}$	&	3.490 	&	36.10 	&	31.96 	&	B0	&	\citet{Kargaltsev2012}\\
J1732$-$3131	&	2.38$\times 10^{12}$	&	1.97$\times 10^{-1}$	&	2.80$\times 10^{-14}$	&	0.640 	&	35.16 	&	30.45 	&	B0	&	\citet{Ray2011}\\
J1740$+$1000	&	1.85$\times 10^{12}$	&	1.54$\times 10^{-1}$	&	2.15$\times 10^{-14}$	&	1.230 	&	35.37 	&	31.88 	&	B0	&	\citet{Kargaltsev2012}\\
J1741$-$2054	&	2.69$\times 10^{12}$	&	4.14$\times 10^{-1}$	&	1.70$\times 10^{-14}$	&	0.300 	&	33.98 	&	31.15 	&	B0	&	\citet{2014ApJ...790...51M}\\
J1747$-$2809	&	2.89$\times 10^{12}$	&	5.22$\times 10^{-2}$	&	1.56$\times 10^{-13}$	&	8.150 	&	37.64 	&	33.45 	&	B0	&	\citet{Porquet2003}\\
J1747$-$2958	&	2.49$\times 10^{12}$	&	9.88$\times 10^{-2}$	&	6.12$\times 10^{-14}$	&	2.520 	&	36.40 	&	33.28 	&	B0	&	\citet{2004ApJ...616..383G}\\
J1809$-$1917	&	1.47$\times 10^{12}$	&	8.27$\times 10^{-2}$	&	2.55$\times 10^{-14}$	&	3.270 	&	36.25 	&	32.13 	&	B0	&	\citet{2007ApJ...670..655K}\\
J1833$-$1034	&	3.58$\times 10^{12}$	&	6.19$\times 10^{-2}$	&	2.02$\times 10^{-13}$	&	4.100 	&	37.53 	&	34.20 	&	B0	&	\citet{2010ApJ...724..572M}\\
J1907$+$0602	&	3.08$\times 10^{12}$	&	1.07$\times 10^{-1}$	&	8.69$\times 10^{-14}$	&	2.580 	&	36.45 	&	31.79 	&	B0	&	\citet{Abdo2010}\\
J2021$+$3651	&	3.19$\times 10^{12}$	&	1.04$\times 10^{-1}$	&	9.57$\times 10^{-14}$	&	1.800 	&	36.53 	&	32.35 	&	B0	&	\citet{2004ApJ...612..389H}\\
J2022$+$3842	&	2.07$\times 10^{12}$	&	4.86$\times 10^{-2}$	&	8.61$\times 10^{-14}$	&	10.000 	&	37.47 	&	34.20 	&	B0	&	\citet{2014ApJ...790..103A}\\
J2043$+$2740	&	3.54$\times 10^{11}$	&	9.61$\times 10^{-2}$	&	1.27$\times 10^{-15}$	&	1.480 	&	34.75 	&	30.76 	&	B0	&	\citet{2013ApJS..208...17A}\\
J2229$+$6114	&	2.04$\times 10^{12}$	&	5.16$\times 10^{-2}$	&	7.83$\times 10^{-14}$	&	3.000 	&	37.35 	&	32.98 	&	B0	&	\citet{KP08}\\
\hline
B1509$-$58	&	1.54$\times 10^{13}$	&	1.51$\times 10^{-1}$	&	1.53$\times 10^{-12}$	&	4.400 	&	37.24 	&	35.13 	&	HB	&	\citet{KP08}\\
B1916$+$14	&	1.60$\times 10^{13}$	&	1.18	&	2.12$\times 10^{-13}$	&	1.300 	&	33.71 	&	30.98 	&	HB	&	\citet{Zhu2009}\\
J0007$+$7303$^{b}$	&	1.08$\times 10^{13}$	&	3.16$\times 10^{-1}$	&	3.60$\times 10^{-13}$	&	1.400 	&	35.65 	&	31.19 	&	HB	&	\citet{2010ApJ...725L...6C}\\
J0726$-$2612	&	3.21$\times 10^{13}$	&	3.44	&	2.93$\times 10^{-13}$	&	2.900 	&	32.45 	&	33.24 	&	HB	&	\citet{Speagle2011}\\
J1119$-$6127	&	4.10$\times 10^{13}$	&	4.08$\times 10^{-1}$	&	4.02$\times 10^{-12}$	&	8.400 	&	36.37 	&	33.34 	&	HB	&	\citet{2012ApJ...761...65N}\\
J1119$-$6127	&	4.10$\times 10^{13}$	&	4.08$\times 10^{-1}$	&	4.02$\times 10^{-12}$	&	8.400 	&	36.37 	&	35.54 	&	HB$^{c}$	&	\citet{Archibald2016}\\
J1124$-$5916	&	1.02$\times 10^{13}$	&	1.35$\times 10^{-1}$	&	7.53$\times 10^{-13}$	&	5.000 	&	37.08 	&	33.31 	&	HB	&	\citet{2001ApJ...559L.153H}\\
J1640$-$4631$^{b}$	&	1.44$\times 10^{13}$	&	2.06$\times 10^{-1}$	&	9.75$\times 10^{-13}$	&	12.750 	&	36.64 	&	34.56 	&	HB	&	\citet{2014ApJ...788..155G}\\
J1718$-$3718	&	7.46$\times 10^{13}$	&	3.38	&	1.61$\times 10^{-12}$	&	3.920 	&	33.22 	&	32.45 	&	HB	&	\citet{Zhu2011}\\
J1734$-$3333	&	5.24$\times 10^{13}$	&	1.17	&	2.28$\times 10^{-12}$	&	4.460 	&	34.75 	&	32.02 	&	HB	&	\citet{Olausen2013}\\
J1819$-$1458	&	5.01$\times 10^{13}$	&	4.26	&	5.75$\times 10^{-13}$	&	3.300 	&	32.47 	&	33.31 	&	HB	&	\citet{McLaughlin2007}\\
J1846$-$0258$^{b}$	&	4.89$\times 10^{13}$	&	3.27$\times 10^{-1}$	&	7.11$\times 10^{-12}$	&	5.800 	&	36.91 	&	34.60 	&	HB	&	\citet{2008ApJ...686..508N}\\
J1846$-$0258$^{b}$	&	4.89$\times 10^{13}$	&	3.27$\times 10^{-1}$	&	7.11$\times 10^{-12}$	&	5.800 	&	36.91 	&	35.26 	&	HB$^{c}$	&	\citet{2008ApJ...686..508N}\\
J1930$+$1852	&	1.03$\times 10^{13}$	&	1.37$\times 10^{-1}$	&	7.52$\times 10^{-13}$	&	7.000 	&	37.06 	&	34.68 	&	HB	&	\citet{2007ApJ...663..315L}\\
\hline
J0501$+$4516	&	1.85$\times 10^{14}$	&	5.76	&	5.82$\times 10^{-12}$	&	2.200 	&	33.08 	&	34.07 	&	Mag	&	\citet{2014MNRAS.438.3291C}\\
J1050$-$5953	&	5.02$\times 10^{14}$	&	6.45	&	3.81$\times 10^{-11}$	&	9.000 	&	33.75 	&	34.89 	&	Mag	&	\citet{2008ApJ...677..503T}\\
J1622$-$4950	&	2.75$\times 10^{14}$	&	4.33	&	1.70$\times 10^{-11}$	&	5.570 	&	33.92 	&	32.60 	&	Mag	&	\citet{2012ApJ...751...53A}\\
J1708$-$4008	&	4.70$\times 10^{14}$	&	1.10$\times 10^{1}$	&	1.96$\times 10^{-11}$	&	3.800 	&	32.76 	&	35.19 	&	Mag	&	\citet{2007ApandSS.308..505R}\\
J1714$-$3810	&	4.81$\times 10^{14}$	&	3.82	&	5.87$\times 10^{-11}$	&	13.200 	&	34.62 	&	34.79 	&	Mag	&	\citet{2010PASJ...62L..33S}\\
J1809$-$1943	&	1.27$\times 10^{14}$	&	5.54	&	2.83$\times 10^{-12}$	&	3.600 	&	32.82 	&	33.92 	&	Mag	&	\citet{2004ApJ...605..368G}\\
J1856$+$0245	&	2.27$\times 10^{12}$	&	8.09$\times 10^{-2}$	&	6.21$\times 10^{-14}$	&	6.320 	&	36.67 	&	33.19 	&	Mag	&	\citet{Rousseau2012}\\
J2301$+$5852	&	5.81$\times 10^{13}$	&	6.98	&	4.71$\times 10^{-13}$	&	3.300 	&	31.74 	&	34.64 	&	Mag	&	\citet{2008ApJ...686..520Z}\\
SGR1627$-$41	&	2.25$\times 10^{14}$	&	2.59	&	1.90$\times 10^{-11}$	&	11.000 	&	34.63 	&	33.66 	&	Mag	&	\citet{2008MNRAS.390L..34E}\\
\hline
RXJ0420.0$-$5022	&	9.95$\times 10^{12}$	&	3.45	&	2.80$\times 10^{-14}$	&	0.340 	&	31.43 	&	30.33 	&	XINS	&	\citet{vigano2013}\\
RXJ0720.4$-$3125	&	2.44$\times 10^{13}$	&	8.39	&	6.90$\times 10^{-14}$	&	0.290 	&	30.66 	&	32.09 	&	XINS	&	\citet{vigano2013}\\
RXJ0806.4$-$4123	&	2.54$\times 10^{13}$	&	1.14$\times 10^{1}$	&	5.50$\times 10^{-14}$	&	0.250 	&	30.17 	&	31.13 	&	XINS	&	\citet{vigano2013}\\
RXJ1308.6$+$2127	&	3.41$\times 10^{13}$	&	1.03$\times 10^{1}$	&	1.10$\times 10^{-13}$	&	0.500 	&	30.60 	&	32.21 	&	XINS	&	\citet{vigano2013}\\
RXJ1605.3$+$3249	&	7.46$\times 10^{13}$	&	3.39	&	1.60$\times 10^{-12}$	&	0.350 	&	33.21 	&	31.92 	&	XINS	&	\citet{vigano2013}\\
RXJ1856.5$-$3754	&	1.47$\times 10^{13}$	&	7.06	&	3.00$\times 10^{-14}$	&	0.120 	&	30.53 	&	31.15 	&	XINS	&	\citet{vigano2013}\\
RXJ2143.0$+$0654	&	1.99$\times 10^{13}$	&	9.43	&	4.10$\times 10^{-14}$	&	0.430 	&	30.29 	&	31.92 	&	XINS	&	\citet{vigano2013}\\
\hline\hline
\multicolumn{9}{l}{$^a$Neutron Star population and group. B0, HB, XINS, Mag listed in the type column are PSRs with inferred dipole magnetic field}\\
\multicolumn{9}{l}{~~${\it B_{\rm d}} < 10^{13}$G and ${\it F_{\rm rot}}={\it L_{\rm rot}}/4\pi D^{2} > 1\times10^{-11}$, high-B RPPs with inferred dipole magnetic field ${\it B_{\rm d}} \lesssim 10^{13}$G and magnetar population.}\\
\multicolumn{9}{l}{$^b$Radio quiet high-B RPPs.}\\
\multicolumn{9}{l}{$^c$High-B RPPs at Burst.}\\
\end{longtable}

\bsp     
\label{lastpage}
\end{document}